\documentclass[letter]{aa} 
\usepackage{graphicx}
\usepackage{natbib}
\usepackage{epsfig}
\def\ut#1{\mathop{\vtop{\ialign{##\crcr
     $\hfil\displaystyle{#1}\hfil$\crcr\noalign
     {\kern1pt\nointerlineskip}\hbox{$\hfil\sim\hfil$}\crcr
     \noalign{\kern1pt}}}}}

\def\undersymbol#1#2{\mathop{\vtop{\ialign{##\crcr
     $\hfil\displaystyle{#2}\hfil$\crcr\noalign
     {\kern1pt\nointerlineskip}\hbox{$\hfil#1\hfil$}\crcr
     \noalign{\kern1pt}}}}}
\def\arcsec{^{\prime\prime}}
\def\arcmin{^{\prime}}
\def\degr{^0}

\usepackage{graphicx}

\begin{document}

\title{Planck revealed bulk motion of Centaurus A lobes}
       \author{F. De Paolis\inst{1,2}, V.G. Gurzadyan\inst{3}, A.A. Nucita\inst{1,2}, G. Ingrosso\inst{1,2},  A.L. Kashin\inst{3}, H.G. Khachatryan\inst{3}, S. Mirzoyan\inst{3}, G. Yegorian\inst{3}, Ph. Jetzer\inst{4}, A. Qadir\inst{5} \and D. Vetrugno\inst{6}}
              \institute{Dipartimento di Matematica e  Fisica ``E. De Giorgi'', Universit\`a del Salento, Via per Arnesano, I-73100, Lecce, Italy  
              \and INFN, Sez. di Lecce, Via per Arnesano, I-73100, Lecce, Italy
              \and Center for Cosmology and Astrophysics, Alikhanian National Laboratory and and Yerevan State
University, Yerevan, Armenia
\and
Physik-Institut, Universit\"at
Z\"urich, Winterthurerstrasse 190, 8057 Z\"urich, Switzerland
\and
School of Natural Sciences,
National University of Sciences and Technology, Islamabad,
Pakistan
\and 
Department of Physics, University of Trento, I-38123 Povo, Trento, Italy and 
TIFPA/INFN, I-38123 Povo,  Italy
}

   \offprints{F. De Paolis, \email{francesco.depaolis@le.infn.it}}
   \date{Submitted: XXX; Accepted: XXX}

 \abstract{Planck data towards the active galaxy Centaurus A are analyzed in the 70, 100 and 143 GHz bands. We find a  temperature asymmetry of the northern radio lobe with respect to the southern one that clearly extends at least up to $5\degr$ from the Cen A center and diminishes towards the outer regions of the lobes. That transparent parameter - the temperature asymmetry - thus has to carry a principal information, i.e. indication on the line-of-sight bulk motion of the lobes, while the increase of that asymmetry at smaller radii reveals the differential dynamics of the lobes as expected at ejections from the center.}


   \authorrunning{De Paolis et al.}
   \titlerunning{Planck revealed bulk motion of Centaurus A lobes}
   \maketitle
%

\section{Introduction}
Centaurus A (and its parent galaxy NGC 5128) is a radio galaxy and represents the closest AGN to us, being at a distance of $3.8\pm 0.1$ Mpc \citep{Harrisetal2010}.
Its jet is clearly visible both in radio and X-rays \footnote{In the radio band it is classified as a Fanaroff-Riley type I low luminosity radio galaxy, as a Seyfert 2 object in the visible and  a ``misdirected''  BL Lac type AGN at high-energy.} and since its discovery \citep{Bolton1948} Cen A has been extensively studied over the entire range of the electromagnetic spectrum (for a review see \citealt{Israel1998}) with a sensitivity and spatial resolution impossible for other active galaxies. It is an extended, morphologically complex (a detailed description of the radio morphology may be found a.g. in \citealt{Burns1983,Meier1989}) and fairly symmetric source exhibiting two giant lobes: the northern one (GLN) and the southern one (GLS), spanning in declination between approximately $-38\degr$ and $-48\degr$ (the coordinates of the Cen A center are R.A. (J2000)= $13^{\rm h} 25^{\rm m} 27.6152^{\rm s}$, Dec. (J2000)=$-43\degr 0.1\arcmin 08.805\arcsec$). We note that an angle of about $10\degr$ on the sky means a physical size at the Cen A distance of $\simeq 600$ kpc in projection and that the redshift of Cen A is z=0.01825, corresponding to a recession velocity of about 540 km s$^{-1}$.

The elliptical (S0) galaxy NGC 5128 is an example of the family of ellipticals that have an absorbing band of gas and dust projected across the stellar body. 
The center of this system harbors a supermassive black hole with mass  about  $10^7-10^8$ M$_{\odot}$ (see e.g. \citealt{Silge2005,Marconi2006,Neumayer2010}) which powers two jets, two inner lobes (with size about a few arcmin each)  and the two outer giant lobes as well. The GLN  and the northern jet are likely tilted towards the observer and the GLN is thought to be closer to us than the GLS. Indeed, the jets appear clearly in the radio band and are obviously shooting out of Centaurus A, with the radio emission becoming more diffuse at greater distances from the  center of the galaxy. The jets consist of a plasma state, i.e. a high-temperature stream of matter.
The jets are also clearly observed in X-rays (see e.g. \citealt{Karovska2002} and references therein). The most prominent feature is the jet extending for about $8$ kpc towards the northeast (upper left in the sky) while a  less prominent jet extends towards the southwest. The apparent brightness difference between the jets and the proper motion asymmetries of both the jets and the inner lobes \citep{Tingay2001} are thought to be due to the viewing geometry: the first jet is moving towards us, while the second is moving away (see e.g. \citealt{Burns1983} and \citealt{Israel1998}).  This was also suggested by the Faraday depolarization analysis of the southern inner lobe \citep{Clarke1992}.

Motivated by the discussion above and by the unique tool that is provided by  data in the microwave region of the electromagnetic spectrum to probe the large-scale temperature asymmetries towards nearby astronomical systems,
in this Letter we study the Cen A system by using {\it Planck} data following the same approach adopted in \cite{depaolis2011,depaolis2014} for the M31 galaxy.

\section{Planck data analysis and results}
\begin{figure}[h!]
 \centering
  \includegraphics[width=0.48\textwidth]{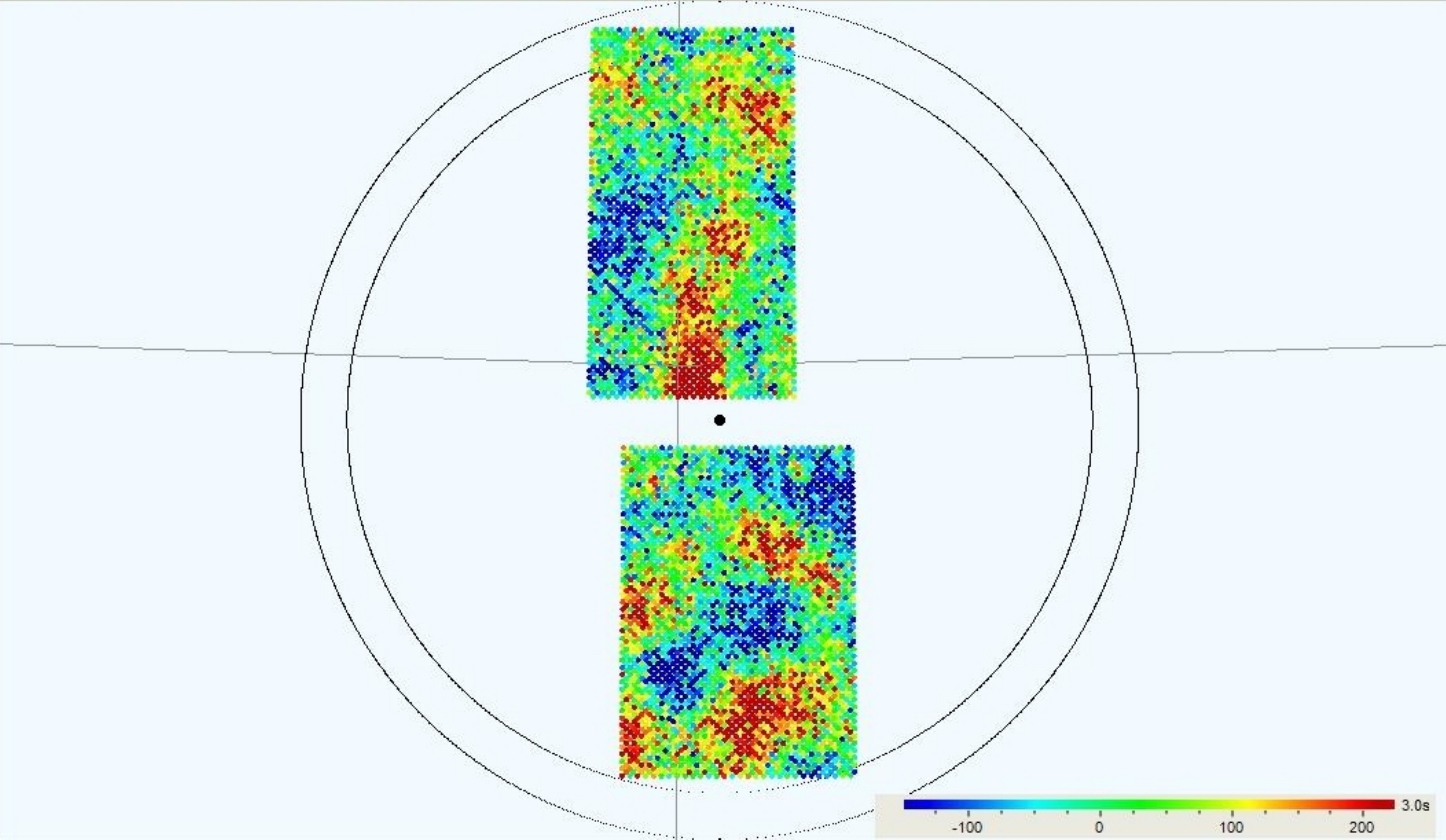}
 \caption{The {\it Planck} fields towards Cen A  galaxy  in which our analysis is performed are shown. The radius of the outer circle is of $5\degr$ about the NGC 5128 center (with Galactic coordinates l=$309.52\degr$, $b=19.42\degr$). The northern field (with 10792 pixels), corresponding to the GLN radio lobe, has Galactic  coordinates $308.65\degr\leq l\leq 311.03$ and $19.64\degr\leq b \leq 23.92\degr$ while those of the southern field (which has 11011 pixels), corresponding to the GLS,  are $307.98\degr\leq l\leq 310.64$ and $15.58\degr\leq b \leq 19.14\degr$.} \label{fig1}
 \end{figure}

We have considered {\it Planck} 2015 release data \citep{planck2015a} in the bands at 70 GHz detected by the LFI instrument, and in the bands at 100 and 143 GHz detected by the HFI instrument (for a review on {\it Planck} results and instruments characteristics we refer to e.g.  \citealt{burigana2013}). 
The resolution in these  {\it Planck} bands are $13\arcmin$, $9.6\arcmin$ and $7.1\arcmin$ (in terms of FHWM) at 70, 100 and 143 GHz, respectively,
 and $N_{side}$=2048 for CMB temperature \citep{planck2015b}.
The sensitivity, angular resolution
and frequency coverage of {\it Planck}  make it a powerful instrument for cosmology as well as galactic and extragalactic astrophysics \citep{planck2015a}.
In order to reveal and study the Cen A giant lobes GLN and GLS in microwaves we have divided the Cen A sky field in two parts as shown in Fig. \ref{fig1} removing the innermost part (about $20\arcmin$) of Cen A corresponding to the NGC 5128 galaxy and the innermost radio lobes.

The mean
temperature excess $T_m$ in $\mu$K in each
region was obtained in each {\it Planck} band at 70, 100 and 143 GHz  with
the corresponding standard error \footnote{The standard error has been 
calculated as the standard deviation of the excess temperature
distribution divided by the square root of the pixel number in
each region. We have verified that within the errors, the sigma
values calculated in that way are consistent with those evaluated
by using the covariance matrix obtained by a best-fitting
procedure with a Gaussian to the same distribution.}.

\begin{figure}[h!]
 \centering
    \includegraphics[width=0.48\textwidth]{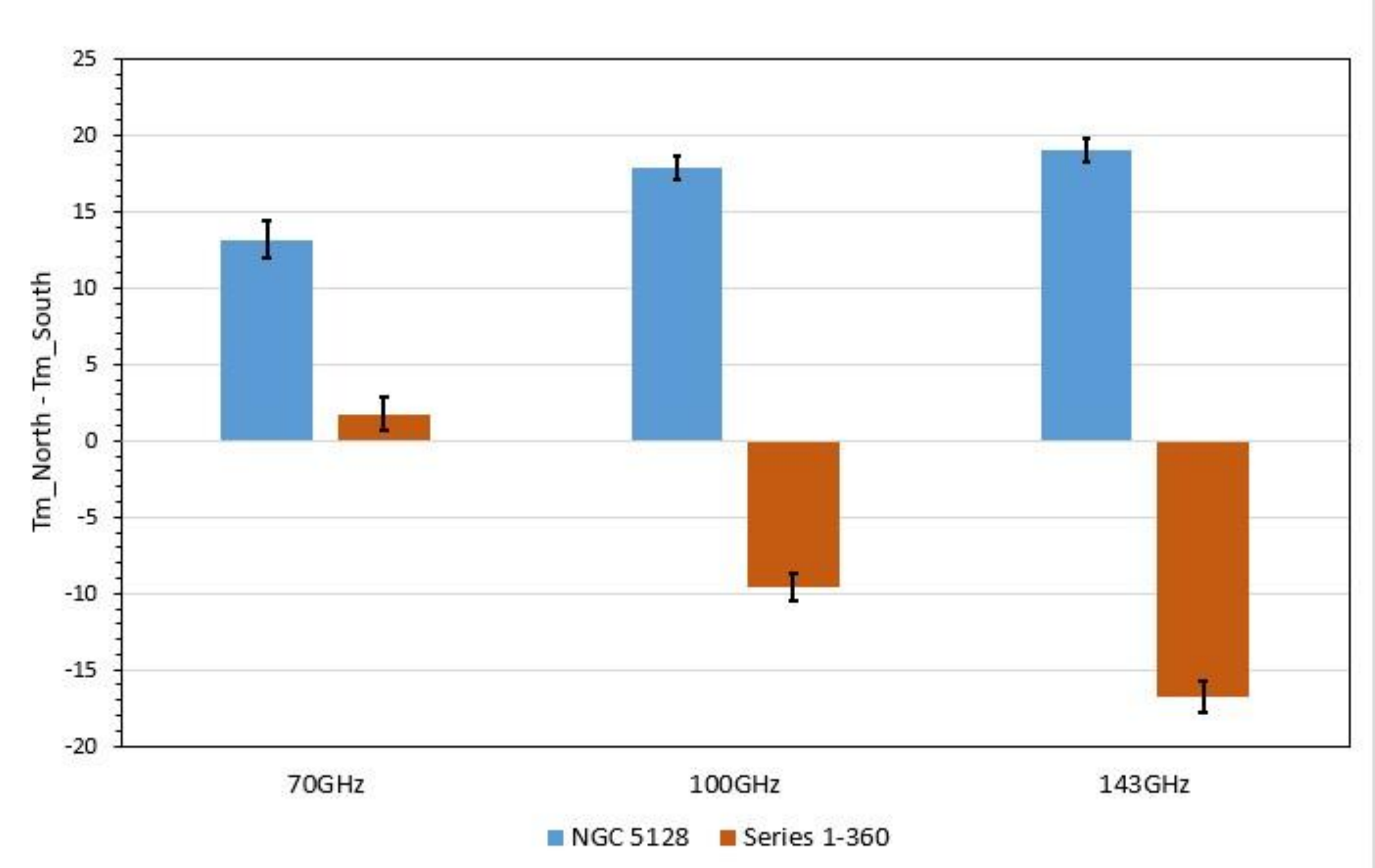}
 \caption{The excess temperature in $\mu$K (in blue) of the northern Cen A lobe with respect to the southern one are given in the 70 GHz, 100 GHz and 143 GHz {\it Planck} data. In brown we give the temperature excess of 360 control regions equally spaced at one degree distance to each other in Galactic longitude and at the same latitude as NGC 5128. The standard errors  are also shown.} \label{fig2}
 \end{figure}
 \begin{figure}[h!]
 \centering
    \includegraphics[width=0.48\textwidth]{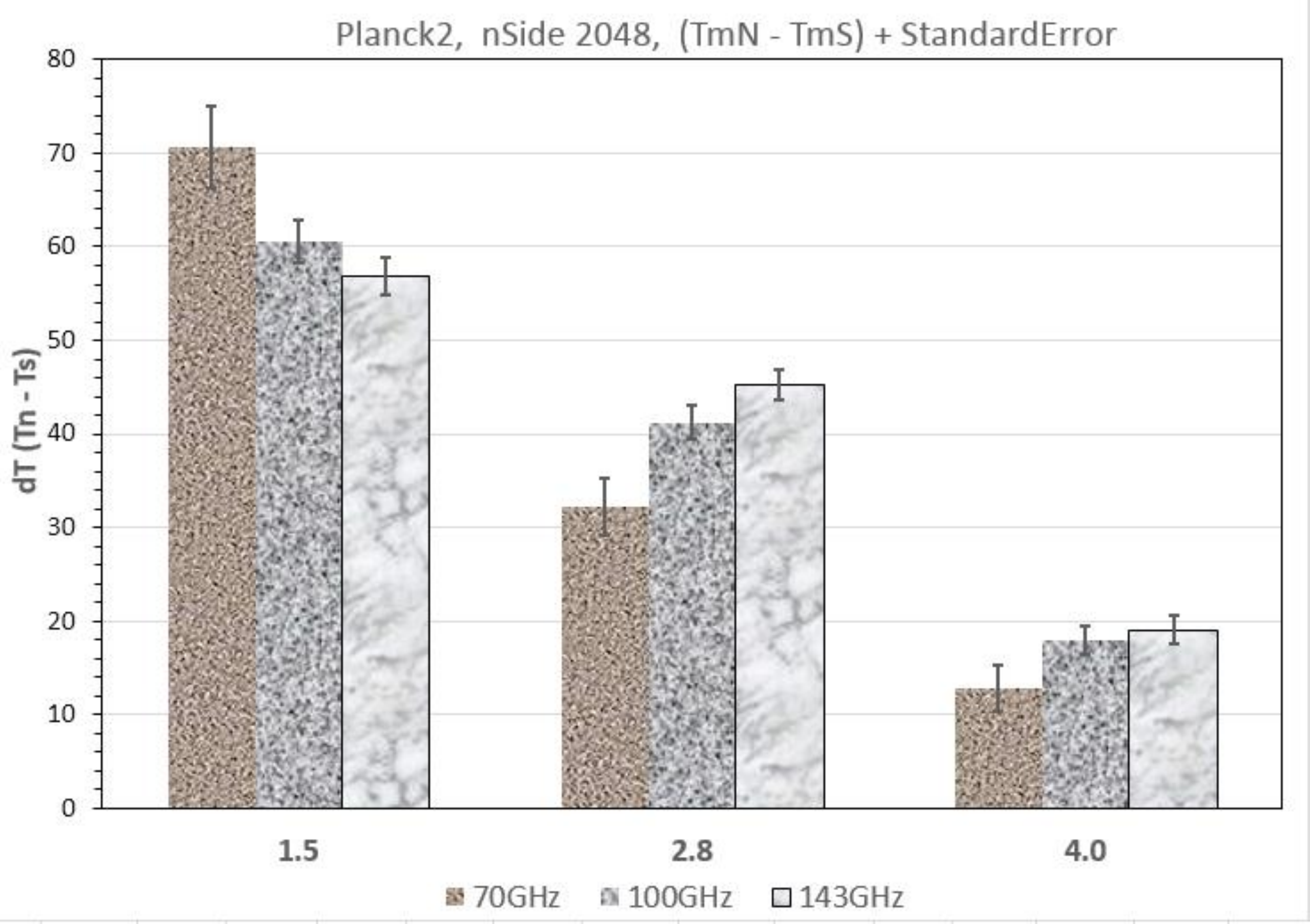}
 \caption{The excess temperature in $\mu$K  (with the standard errors) of the northern Cen A lobe with respect to the southern one are given in the 70 GHz, 100 GHz and 143 GHz {\it Planck} data up $1.5\degr$ ($\simeq 92$ kpc), $2.8\degr$ ($\simeq 171$ kpc) and $4.0\degr$ ($\simeq 245$ kpc) from the galaxy center.} \label{fig3}
 \end{figure}

The results obtained for the temperature difference between the GLN and the GLS temperature in each of the three {\it Planck} bands are shown in Fig. \ref{fig2}.
As one can see (in blue), the GLN region is hotter than the GLS region by about 13 $\mu$K at 70 GHz, 17 $\mu$K at 100 GHz and 20 $\mu$K at 143 GHz (in blue in the histogram).  This results, most likely, from a Doppler-induced effect related to the bulk velocity of the jet powering the Cen A radio lobes and/or to their rotation with respect to an axis directed along the East-West direction.

To test whether the temperature asymmetry we see is real or can be explained as a random fluctuation of the CMB signal (which  is very patchy) we  consider 360 control field regions with the same shape as the GLN and GLS regions and at the same latitude, but at  $1\degr$ longitude from each other. For each region we determined the excess temperature profile and calculated the average profile and corresponding standard deviation. The results obtained are  shown in brown in the histogram in Fig. \ref{fig2}. As one can see, the temperature asymmetry in the 360 control fields at 100 and 143 GHz is the opposite of that towards Cen A with the southern field generally hotter than the northern one (in the 70 GHz band the temperature asymmetry of the control fields is substantially smaller with respect to that towards the Cen A lobes). This trend for the 360 control fields is easily understandable, since the Galactic latitude of the southern lobes is $14-19\degr$ (the center of Cen A is at latitude $b=19.473\degr$) and at these latitudes the foregrounds due to our Galactic disk is non-negligible. Indeed, since the  Cen A giant lobes happen to be aligned in the sky almost orthogonal to the Milky Way, the southern lobe has to be more contaminated by the disk's radiation than the northern one.  We also stress that the procedure we followed to test the observed effect, that is of considering the 360 control regions, is more reliable than simulating the CMB sky maps in each band considered. While the latter methodology of generating sky maps to estimate the error bars is mandatory  dealing with the whole sky (as in cosmological studies), in our case we are considering only rather small regions of the {\it Planck} sky maps and the adopted procedure is more reliable since it avoids the simulation ambiguities. 

In Fig. \ref{fig3} the analysis of the temperature asymmetry between the GLN and GLS of Cen A within three different galactocentric radii ($1.5\degr$, $2.8\degr$ and $4.0\degr$, corresponding to about 92 kpc, 171 kpc and 245 kpc, respectively) is presented. It is clear that the temperature asymmetry diminishes as increasing the field region area. The radial dependence of the temperature asymmetry in the three  {\it Planck} bands considered might be indicative of  different emission mechanisms in the microwaves at  different galactocentric distances, even if the quality of the present data cannot allow us to draw a definitive conclusion in this respect.

\section{Conclusions}

As discussed in the previous section, we have considered {\it Planck} 2015 release data in the bands at 70, 100 and 143 GHz  and detected a temperature asymmetry between the GLN and the GLS of the Cen A system. Notice that  no detection of the Cen A giant lobes was present until now in the microwaves at wavelengths higher than 60 GHz \citep{Hardcastle2009}. In particular, the GLN appears substantially hotter than the GLS up to a galactocentric distance of about $5\degr$. 
The temperature  asymmetry is present in all the bands considered and decreases from the innermost region (being about 70 $\mu$K at 70 GHz within  $1.5\degr$ from the Cen A center) to about 14 $\mu$K at 70 GHz within $4\degr$ from the galaxy center. What we find seems to confirm what is known from radio observations about the emission direction and motion of the Cen A jet and inner northern and southern lobes.

Since the  Cen A giant lobes happen to be aligned in the sky almost orthogonal to the Galactic disk, as mentioned, the southern lobe has to be more contaminated by the disk's radiation than the northern one, giving an effect opposite to that observed towards the Cen A lobes. This fact is reflected in the behavior of the mean temperature asymmetry, including the band dependence, of the 360 control regions. 
Hence, in view of Galactic contamination, the genuine Cen A temperature contrast has to be even stronger. Moreover, since we are not dealing with the absolute but with the mean temperature differences  only, the role of the various noise sources is vanishing. This is similar to the case of the CMB dipole, where the temperature difference indicates the motion of the observer.   

In general, the observation of the Cen A temperature asymmetry may be explained by one (or more) of the following astrophysical emission mechanisms: ($i$) free-free emission, ($ii$) synchrotron emission, ($iii$) anomalous microwave emission (AME) from dust grains, ($iv$) the kinetic
Sunyaev-Zel'dovich (SZ) effect, and ($v$) cold gas clouds populating the outer regions of Cen A (as first proposed, in the context of the M31 halo, by \citealt{depaolis1995}). A detailed study of what each of these five possible causes might contribute, using all the {\it Planck} bands to constrain the model parameters and the relative weight of these five models, will be published elsewhere. Here, we only note that effects $(i)-(iii)$ are strongly wavelength dependent at microwave frequencies (see, e.g. \cite{bennett2003,planck2013}, while $(iv)$ and $(v)$ are almost independent of the observation band in the microwave regime. 
AME (item $iii$) has been observed in
various interstellar environments, in particular in the diffuse ISM \citep{miville2008} and in dark
clouds \citep{Watson2005}, and might
play a role also in galactic halo environments, provided dust grains are present.\footnote{AME should give an effect strongly frequency dependent in the CMB domain, too.}

As a matter of fact, and irrespective of the physical emission mechanisms, the detected mean temperature asymmetry has to indicate the line-of-sight motion of the lobes.

In the viewing geometry, a GLN hotter than the GLS will be due to the direction of the powering jets, as also suggested  by observations in other wavelengths (see e.g. the discussion in the Introduction), also a Doppler induced effect due to the rotation of the giant radio lobes with respect to an axis directed along the East-West direction would contribute, an effect similar to that observed towards the halo of the M31 galaxy 
\citep{depaolis2011, depaolis2014}. Although the resolution of this issue is not the aim of the present Letter,  we shall give some possible hints in the following.
The rotation of NGC 5128 and its halo has been investigated by studying the velocity distribution of more than 400 planetary nebulae within a galactocentric distance of about 
$20\arcmin$ ($\simeq 20$ kpc). In particular, \cite{Hui1995} found that the NGC 5128 rotation axis is offset from its photometric minor axis by about $39\degr$. It was also found that the planetary nebulae ordered motions become more important with respect to their random motions at larger galactocentric radii, with the rotation component reaching  about 100 km s$^{-1}$ and 50 km s$^{-1}$ along the photometric major and minor axes, respectively.

The kinematics of the globular clusters around Cen A gives a similar indication, in particular with the metal rich ones showing both major and minor axis rotation \citep{Hui1995}.
This was also confirmed by a more recent analysis by \cite{Woodley2010} who considered a sample of 605 globular clusters extending up to a galactocentric distance of about $45\arcmin$. It was found that the metal rich globular clusters are rotating with an ordered speed of 43 km s$^{-1}$ while the metal poor ones have a very mild rotation signature of $\simeq 26$ km s$^{-1}$.
We have independently analyzed the radial velocity measured in the globular cluster sample in \cite{Woodley2010} and \cite{Peng2004} and, after removing the innermost globular clusters (within a galactocentric radius of about $4\arcmin$) we found that those in the northern lobe (with declination $\delta \geq -42.57\degr$) have radial velocity $470\pm143$ km s$^{-1}$ while those in the southern lobe (with $\delta \leq -43.04\degr$) have  radial velocity $580\pm148$ km s$^{-1}$. This  seems to indicate that there is a regular rotation component at least of the innermost side of the Cen A lobes. 

We would also like to mention that such mean temperature asymmetry method first applied to study the M31 galaxy \citep{depaolis2011, depaolis2014} and applied here to the giant radio lobes of Cen A can become a conventional tool for studying of internal motions, especially towards nearby galaxies, in the microwaves. As for the case of the SZ effect or e.g. for the Kolmogorov stochasticity parameter 
(\cite{Gurzadyan2009,Gurzadyan2014} and references therein), software for an automatic analysis by correlating galaxy surveys and CMB data may be developed. 

The detected Cen A temperature asymmetry and especially its increase at small radii outline the picture of continuous ejections from a center which is rotating, and upon the increase of the size of the lobes, the effect of differential rotation becomes noticeable.
Moreover, as one can see from Fig. 3, the temperature asymmetry within $1.5\degr$ tends to decreases from the 70 GHz band to the 143 GHz band, while it goes in the opposite directions at outer radii. This trend, if confirmed, seems to indicate that the dominant emission mechanism at CMB frequencies changes somewhere in between $1.5 \degr$ and $2.8\degr$, that is between about 92 kpc and 170 kpc.

It goes without saying that understanding the reason for the Cen A temperature asymmetry, that is if it is dominantly due to the ejection direction of the jets or to the rotation of the inner and outer lobes is of particular importance since it can throw light on the formation history and timescale evolution of this system and may allow us  to  get closer to the solution of the many unresolved questions about Cen A's giant radio lobes (see e.g.  \citealt{Eilek2014}). To this aim, we suggest to perform an analysis in the radio band, at 21 cm, similar to that done by 
\cite{Chemin} and \cite{Corbelli} for the disk of the M31 galaxy with the aim of tracking a radial velocity map of the outer regions of the Cen A system.

\begin{acknowledgements}
{We acknowledge the use of {\it Planck}'s data in the Legacy Archive for
Microwave Background Data Analysis (LAMBDA) and HEALPix
\citep{gorski/etal:2005} package. We thank L. Chemin for fruitful discussion. FDP, AAN and GI acknowledge the support by the INFN
project TAsP and PJ acknowledges support from the
Swiss National Science Foundation.}
\end{acknowledgements}



\begin{thebibliography}{99}
\bibitem[\protect\citeauthoryear{Bennett}{2003}]{bennett2003} 
Bennett, C.L. 2003,  ApJ, 148, 97
\bibitem[\protect\citeauthoryear{Bolton}{1948}]{Bolton1948} 
Bolton, J.G. 1948, Nature, 161, 141
\bibitem[\protect\citeauthoryear{Burigana et al.}{2013}]{burigana2013}
Burigana, C., Davies, R.D., De Bernardis, P. et al. 2013, IJMPD, 22,  id. 1330011
\bibitem[\protect\citeauthoryear{Burns et al.}{1983}]{Burns1983} 
 Burns, J.O., Feigelson, E.D. \& Schreier, E.J. 1983, ApJ, 273, 128
 \bibitem[\protect\citeauthoryear{Chemin et al.}{2009}]{Chemin}
Chemin, L., Carignan, C. \& Foster, T. 2009, ApJ, 705, 1395
 \bibitem[\protect\citeauthoryear{Clarke et al.}{1992}]{Clarke1992} 
 Clarke, D. A., Burns, J. O., and Norman, M. L. 1992, ApJ, 395, 444
 \bibitem[\protect\citeauthoryear{Corbelli et al.}{2010}]{Corbelli} Corbelli, E., Lorenzoni, S., Walterbor, R.  et al. 2010,
A\&A,  511, id. A89
 \bibitem[\protect\citeauthoryear{De Paolis et al.}{1995}]{depaolis1995} De Paolis, F., Ingrosso, G., Jetzer, Ph. et al. 1995,
A\&A, 299, 647
 \bibitem[\protect\citeauthoryear{De Paolis et al.}{2011}]{depaolis2011} De Paolis, F., Gurzadyan, V.G., Ingrosso, G.  et al. 2011,
A\&A, 534, id.  L8
\bibitem[\protect\citeauthoryear{De Paolis et al.}{2014}]{depaolis2014} De Paolis, F., Gurzadyan, V.G., Nucita, A.A.  et al. 2014,
A\&A, 565, id.  L3
\bibitem[\protect\citeauthoryear{Eilek}{2014}]{Eilek2014}
Eilek, J.A. 2014, New Journal of Physics 16, 045001
\bibitem[{{G{\'o}rski} {et~al.}(2005){G{\'o}rski}, {Hivon}, {Banday},
  {Wandelt}, {Hansen}, {Reinecke}, \& {Bartelmann}}]{gorski/etal:2005}
{G{\'o}rski}, K.~M., {Hivon}, E., {Banday}, A.~J. et al. 2005,  \apj, 622, 759
\bibitem[\protect\citeauthoryear{Gurzadyan et al.}{2009}]{Gurzadyan2009} 
Gurzadyan, V.G., Allahverdyan, A.E., Ghahramanyan, T., et al. 2009, A\&A, 497, 343
\bibitem[\protect\citeauthoryear{Gurzadyan et al.}{2014}]{Gurzadyan2014} 
Gurzadyan, V.G., Kashin, A.L., Khachatryan, H.G. et al. 2014, A\&A, 566, A135 
\bibitem[\protect\citeauthoryear{Hardcastle et al.}{2009}]{Hardcastle2009} 
Hardcastle, M., Cheung, C., Feain, I. and Stawarz, L. 2009, MNRAS, 393, 1041
\bibitem[\protect\citeauthoryear{Harris et al.}{2010}]{Harrisetal2010} 
Harris, G.L.H., Rejkuba, M. and Harris, W.E. 2010, Publications of the Astronomical Society of Australia, 27, 457
\bibitem[\protect\citeauthoryear{Hui et al.}{1995}]{Hui1995} 
Hui, X., Ford, H.C., Freeman, K.C. and Dopita, M.A. 1995, ApJ, 449, 592
\bibitem[\protect\citeauthoryear{Israel}{1998}]{Israel1998} 
Israel, F.P. 1998, AA\&R, 8, 237
\bibitem[\protect\citeauthoryear{Karovska et al.}{2002}]{Karovska2002} 
Karovska, M., Fabbiano, G., Nicastro, F. et al. 2002, ApJ, 577, 114
\bibitem[\protect\citeauthoryear{Marconi et al.}{2006}]{Marconi2006} 
Marconi, A., Pastorini, G., Pacini, F. et al. 2006, A\&A, 448, 921 
\bibitem[\protect\citeauthoryear{Meier et al.}{1989}]{Meier1989} 
Meier, D.L., Jauncey, D.L., Preston, R.A. et al.   1989, AJ, 98,  27
\bibitem[\protect\citeauthoryear{Miville-Desch\^enes et al.}{2008}]{miville2008} 
Miville-Desch\^enes, M.A., Ysard, N., Lavabre, A. et al. 2008, A\&A, 490, 1093
\bibitem[\protect\citeauthoryear{Neff et al.}{2015}]{Neff2015} 
Neff, S. G., Eilek, J.A. and Owen, F.N. 2015, ApJ, 802,  id. 88
\bibitem[\protect\citeauthoryear{Neumayer}{2010}]{Neumayer2010} 
Neumayer, N. 2010, Publications of the Astronomical Society of Australia, 27, 449
\bibitem[\protect\citeauthoryear{Peng et al.}{2004}]{Peng2004} 
Peng, E.W. et al. 2004, ApJS, 150, 367
\bibitem[\protect\citeauthoryear{Planck Collaboration XII}{2013}]{planck2013} 
Planck Collaboration XII, 2014, A\&A, 571,  id.A12
 \bibitem[\protect\citeauthoryear{Planck Collaboration I}{2015}]{planck2015a} 
Planck Collaboration I, 2015, A\&A submitted; arXiv:1502.01582
\bibitem[\protect\citeauthoryear{Planck Collaboration XVI}{2015}]{planck2015b} 
Planck Collaboration XVI, 2015, A\&A submitted; arXiv:1506.07135
\bibitem[\protect\citeauthoryear{Silge et al.}{2005}]{Silge2005} 	
Silge, J.D., Gebhardt, K., Bergmann, M. and Richstone, D. 2005, AJ, 130, 406
\bibitem[\protect\citeauthoryear{Tingay et al.}{2001}]{Tingay2001} 
Tingay, S. J., Preston, R. A., and Jauncey, D. L.  2001, AJ, 122, 1697
\bibitem[\protect\citeauthoryear{Watson et al.}{2005}]{Watson2005} 
Watson, R.A., Rebolo, R., Rubino-Martin, J.A. et al., 2005, ApJ, 624, L89
\bibitem[\protect\citeauthoryear{Woodley et al.}{2010}]{Woodley2010} 
Woodley, K.A., G\'omez, M., Harris, W.E., Geisler, D.  and Harris, G.L.H. 2010, AJ, 139, 1871




\end{thebibliography}
\end{document}